\documentstyle[12pt]{article}

\makeatletter
\def\eqnumsection{\@addtoreset{equation}{section}\def\theequation
{\arabic{section}.\arabic{equation}}}
\makeatother

\title{
\begin{flushright}
{\large Yaroslavl State University\\
        Preprint YARU-HE-96/03} \\[10mm]
\end{flushright}
       {\LARGE\bf Neutrino Radiative Decay} \\ 
       {\LARGE\bf in an External Electromagnetic Field} \\ 
       {\LARGE\bf via Vector Leptoquark}} 

\author{{\Large\bf N.V.~Mikheev, M.L.~Zilberman} \\[2mm]
        {\large\it
             Division of Theoretical Physics, Department of Physics,} \\
        {\large\it
             Yaroslavl State University, Yaroslavl 150000, Russia} \\[4mm]
        {\Large\bf and} \\[4mm]
        {\Large\bf L.A.~Vassilevskaya} \\[2mm]
        {\large\it
             Moscow Lomonosov University, V-952, Moscow 117234, Russia} \\
        {\large\it E-mail: vassilev@yars.free.net, vasilevs@vitep5.itep.ru}}

\date{}

\begin{document}

\maketitle

\eqnumsection

\begin{abstract}
The massive neutrino radiative decay 
$\nu_i \rightarrow \nu_j \gamma$ ($i \ne j$)
is investigated within the 
Minimal Quark-Lepton Symmetry of the Pati-Salam type based on the 
$SU(4)_V \times SU(2)_L \times G_R$ group in an external constant
crossed field ($\vec {\cal E} \perp \vec H$, ${\cal E}=H$).
The matrix element $\Delta M^{(X)}$ and the decay probability 
$W^{(X)}$ due to the leptoquark $X$ contribution are analysed. 
The effect of significant enhancement of the neutrino decay by the
crossed field takes place. It is shown that the leptoquark contribution 
to the decay probability can dominate over the $W$-boson one 
in the case of strong suppression of the lepton mixing
in the framework of the SM.
\end{abstract}

\vspace*{10mm}

\centerline{To be published in Physics Letters B}

\thispagestyle{empty}

\newpage

\large

\section{Introduction}

The Standard Model (SM) successfully describes all the available
experimental data obtained at currently accessible energies. Although
significant progress has been achieved in experimental studies of
electroweak processes, some problems that are important for the
theoretical description of elementary-particle physics remain 
unresolved.
The phenomenon of fermion mixing in charged weak currents
(whose mechanism has yet to be understood) is among the most intriguing
problems of this kind. Mixing in the quark sector is described by the
Cabbibo-Kobayashi-Maskawa  $(3 \times 3)$ unitary matrix $V_{ij}$. 
It is known to a rather high accuracy ~\cite{PDG}, and the information
about mixing parameters is continually refined. It is natural to
expect that a similar mixing phenomenon occurs in the lepton sector
of the SM as well, 
provided that the neutrino mass spectrum is nondegenerate.
It may lead to such interesting phenomena as neutrino 
oscillations~\cite{BP}, rare decays with lepton number violation
of the types $\mu \rightarrow e \gamma$~\cite{Bil}, 
$\mu \rightarrow e \gamma \gamma$~\cite{VGM1}, 
$\nu_i \rightarrow \nu_j \gamma$~\cite{Bil}, 
$\nu_i \rightarrow \nu_j \gamma \gamma$~\cite{Niv}, and other
processes "forbidden" by the law of conservation of lepton number.
Notice that both oscillations and above-mentioned
radiative decays are continuously searched for in 
experiments~\cite{BV,Moh}.

It is necessary to stress that the probabilities of these decays 
are strongly suppressed in the SM due to the well-known 
GIM cancellation by the factor

\begin{equation}
\left (\frac{m_l}{m_W} \right )^4  \ll 1, 
\label{eq:GIM}
\end{equation}

\noindent The small values of lepton masses 
with respect to the $W$-boson mass
lead to the conclusion that processes with such
low probabilities can hardly be observed under laboratory conditions. 
It is worth noting that an additional type of sup\-pres\-sion
due to small mixing angles can appear.
The experimental searches for neutrino oscillations, the main
source of information on the lepton mixing angles, show that the
mixing angles are most likely to be small.
All attempts to find a possible manifestation of lepton
mixing have given up to now negative results.

On the other hand, the continuing interest in neutrinos involving 
processes is due to that the study of their properties may lead to
new physics beyond the SM~\cite{Moh}.
It is conceivable that new symmetries will
successively manifest themselves in experiments with increasing energy
of colliding particles and that $SU(2) \times U(1)$-symmetry of the SM
is merely the first level in the as-yet-unknown hierarchy of symmetries.
In this connection, it is interesting to find the next (in energy)
level in this hierarchy. This brings up the question of what symmetry 
is restored after the electroweak one and of what mass scale follows
$M_W$. There is a possibility of extending the SM in such a way that the
rare radiative lepton decays:

\noindent 1) take place in the absence of lepton mixing in the SM,
 
\noindent 2) are not subjected to the above-mentioned strong suppressions.

\noindent We have in mind the Minimal Quark-Lepton Symmetry of 
the Pati-Salam type based on the 
$SU(4)_V \times SU(2)_L \times G_R$ 
gauge group, in which the lepton
number is interpreted as the fourth color~\cite{PSm}.  
Let us assume that the right-hand symmetry $G_R$ is restored on a
considerably higher mass scale than $SU(4)_V$; therefore, we do not
specify the subgroup $G_R$. 
Fractionally charged color $X$ bosons -- 
leptoquarks -- which are responsible for the quark-lepton transitions, 
are the most exotic objects in this model.
It was shown in~\cite{KM} that a new type of mixing -- mixing in
quark-lepton currents -- must be
considered in this approach. The resulting additional freedom in
the choice of parameters makes it possible to remove the lower bound 
on the vector-leptoquark mass  $M_X$ obtained from rare decays of
$\pi$ and $K$ mesons~\cite{PDG}. The only constraint that does not
depend on mixing arises from the cosmological estimate of the width 
of the decay
$\pi^0 \rightarrow \nu \bar{\nu}$~\cite{Lam}: $M_X > 18$~'eV.

At present the investigation of electroweak processes
in strong external fields are of special interest. In this case,
the method of exact solutions of relativistic wave equations in 
external electromagnetic fields is quite effective and allows one
to go beyond the perturbation theory and predict new phenomena. 
In particular, an external field can induce novel lepton
transitions of the types
$\nu_i \rightarrow \nu_j$ ($i \ne j$)~\cite{BTV},
$\mu \rightarrow e $~\cite{ABV},
forbidden by energy-momentum conservation in vacuum, and influence 
substantially the neutino processes with flavour violation (one-loop
as minimum)~\cite{GMV2,GMV3}.

In the present work we study the massive neutrino radiative decay
$\nu_i \rightarrow \nu_j \gamma$ ($i \ne j$)within the 
Minimal Quark-Lepton Symmetry of the Pati-Salam type based on the 
$SU(4)_V \times SU(2)_L \times G_R$ group in an external constant
crossed field ($\vec {\cal E} \perp \vec H$, ${\cal E}=H$).

\section{The amplitude of $\nu_i \rightarrow \nu_j \gamma$
         in the external crossed field}

\subsection{The quark-lepton interaction Lagrangian}

The Lagrangian describing the interaction of the $up$--type fermion 
with the leptoquark has the form:

\begin{equation}
{\cal L}_X \, = \, \frac{g_s(M_X)}{\sqrt 2} \big [
{\cal U}_{i q}
\big ( \bar \nu_i \gamma_{\alpha} q^c \big ) X^c_{\alpha} + h.c. 
\big ] \, ,
\label{eq:LX}
\end{equation}

\noindent where $c$ is the $SU(3)$ color index, the indices $i$ 
and $q$ correspond to the $up$--fermions: 
the neutrino states $\nu_i$ and quarks $q = u,c,t$ 
are the eigenstates of the mass matrix.
The constant  \, $g_s(M_X)$ \, 
can be expressed in terms of the strong coupling constant
 \, $\alpha_s$ \, at the leptoquark mass scale
$M_X, \quad g_s^2(M_X)/4 \pi = \alpha_s(M_X)$. 
Our knowledge of the properties of the unitary matrix $\cal U$ 
is restricted to the fact that the product of $\cal U$ 
and a similar matrix $\cal D$ in the interaction Lagrangian of
$down$--fermions with leptoquarks equals the standard 
Cabbibo-Kobayashi-Maskawa matrix: $V = {\cal U^+ D}$. 

If the momentum transferred  \, $k^2 \ll M_X^2$, \,  then the 
Lagrangian~(\ref{eq:LX}) leads to an effective four-fermion 
interaction of the
quark-lepton vector currents. By using the Fiertz transformation,
it can be presented as the interaction of the scalar, pseudoscalar, 
vector and axialvector lepton and quark currents:

\begin{eqnarray}
{\cal L}_{eff} & = &  \frac{2 \pi \alpha_S(M_X)}{M_X^2} \;
{\cal U}_{iq} {\cal U}^*_{j q} 
 \, \Bigg{\{ } \,  ( \bar \nu_j \nu_i \big ) \big ( \bar q q \big ) -
 ( \bar \nu_j \gamma_5 \nu_i \big ) ( \bar q \gamma_5 q \big ) 
\nonumber \\
& - &  {\frac{1}{2}} ( \bar \nu_j \gamma_{\alpha} \nu_i ) 
 ( \bar q \gamma_{\alpha} q ) - 
 {\frac{1}{2}} ( \bar \nu_j \gamma_{\alpha} \gamma_5 \nu_i ) 
 ( \bar q \gamma_{\alpha} \gamma_5 q )  \, \Bigg{\} } \; .
\label{eq:LF}
\end{eqnarray}

\noindent In constructing the effective 
Lagrangian of lepton-quark interactions, 
we must take QCD corrections into account, which can be estimated
according to well-known procedures~\cite{VZS}.
In our case QCD corrections are reduces
to the enhancement factor for the scalar and pseudoscalar 
couplings~\cite{Kuz}. 
However, as we shall see later the scalar and pseudoscalar currents 
give small contribution to the ultrarelativistic neutrino decay with
respect to the axial current. By this means taking into account
QCD corrections isn't essential in this case.

\subsection{The amplitude of $\nu_i \rightarrow \nu_j \gamma$
         decay}

The amplitude of the neutrino radiative decay in the one-loop
approximation is described by the Feynman
diagram, represented in Fig.1, where double lines imply the
influence of the external field in the propagators of intermediate
fermions ($f = l, q$; $l = e, \mu, \tau$, $q = u, c, t$). 
The field-induced contribution to the matrix element $\Delta M$ is:

\begin{eqnarray}
\Delta M = \Delta M^{(W)} + \Delta M^{(X)},
\label{eq:SWX}
%\nonumber 
\end{eqnarray}

\noindent where $\Delta M^{(W)}$ and  $\Delta M^{(X)}$ are the 
contributions from $W$-boson and leptoquark $X$, respectively. 
The purpose of this paper is to investigate the leptoquark contribution 
to this process as the $W$-boson one was investigated in detail 
in the uniform magnetic field ~\cite{GMV2} and in the monochromatic
wave field ~\cite{GMV3}. The crossed field limit of the $W$-boson
contribution to the amplitude $\Delta M^{(W)}$ 
can be obtained, for example, from the matrix element 
$\Delta S$ ($i \neq j$)~\cite{GMV3} in the regular way when the wave 
frequency $\omega$ tends to zero with fixed field strengths.

%%%%%%%%%%%%%%%  BEGIN FIGURE  %%%%%%%%%%%%%%%%%%%%%%%%%%%%%%%%%%%

\begin{figure}[tb]

\unitlength=1.00mm
\special{em:linewidth 0.4pt}
\linethickness{0.4pt}

\begin{picture}(60.00,45.00)(-50.00,10.00)

\put(35.00,32.50){\oval(20.00,15.00)[]}
\put(35.00,32.50){\oval(16.00,11.00)[]}
\put(26.00,32.50){\circle*{3.00}}
\put(44.00,32.50){\circle*{2.00}}

\linethickness{0.8pt}

\put(11.00,42.50){\vector(3,-2){9.00}}
\put(26.00,32.50){\line(-3,2){6.00}}
\put(26.00,32.50){\vector(-3,-2){9.00}}
\put(17.00,26.50){\line(-3,-2){6.00}}

\put(36.50,39.00){\line(-3,2){4.01}}
\put(36.50,39.00){\line(-3,-2){4.01}}

\put(32.50,26.00){\line(3,2){4.01}}
\put(32.50,26.00){\line(3,-2){4.01}}

\put(23.00,28.00){\makebox(0,0)[cc]{\large $x$}}
\put(47.00,28.00){\makebox(0,0)[cc]{\large $y$}}
\put(18.00,42.00){\makebox(0,0)[cb]{\large $\nu_i(p)$}}
\put(16.00,23.00){\makebox(0,0)[ct]{\large $\nu_j(p')$}}
\put(34.00,45.00){\makebox(0,0)[cc]{\large $f$}}
\put(34.00,20.00){\makebox(0,0)[cc]{\large $f$}}
\put(55.00,36.00){\makebox(0,0)[cb]{\large $\gamma(k)$}}

%%%%%%%%% Definition for the photon line %%%%%%%%%%%%%%%%%%%%%%

\def\photonatomright{\begin{picture}(3,1.5)(0,0)
                                \put(0,-0.75){\tencircw \symbol{2}}
                                \put(1.5,-0.75){\tencircw \symbol{1}}
                                \put(1.5,0.75){\tencircw \symbol{3}}
                                \put(3,0.75){\tencircw \symbol{0}}
                      \end{picture}
                     }
\def\photonrighthalf{\begin{picture}(30,1.5)(0,0)
                     \multiput(0,0)(3,0){5}{\photonatomright}
                  \end{picture}
                 }

%%%%%%%%%%%%%%%%%%%%%%%%%%%%%%%%%%%%%%%%%%%%%%%%%%%%%%%%%%%%%%%

\put(44.00,32.50){\photonrighthalf}

\end{picture}

\caption{}

\end{figure}

%%%%%%%%%%%%%%%  END FIGURE  %%%%%%%%%%%%%%%%%%%%%%%%%%%%%%%%%%%

The expression for the $\Delta M^{(W)}$ corresponds to Fig.1
with $f = l$ and can be presented in the form:

\begin{eqnarray}
{\Delta M^{(W)}} & = & {e G_{F} \over 4 \sqrt 2 \pi^2 } \,
\sum \limits_{l= e, \mu,\tau} {\cal K}_{il} {\cal K}^*_{j l} \,
\nonumber \\
& \times &
\Bigg \{ \, e (\tilde{F}f^{*}) \,
{(kFFj) \over (kFFk)} \, J_1(\eta_l) 
+  {e \over 8 m^2_l} \,
(F \tilde{f}^{*})(kj) \, J_{2,-1}(\eta_l)  
\label{eq:Amp2} \\
& + & {e^2 \over 24 m^4_l} \left ( (F \tilde{f}^{*}) \, (k\tilde{F}j) \,
J_{3,-1}(\eta_l) + {1 \over 2} \, (Ff^{*}) \, (kFj) \, J_{3,1}(\eta_l) \,
\right ) \Bigg \}, 
\nonumber \\
j_\mu & = & \bar \nu_j (p') \gamma_\mu (1 + \gamma_5) \nu_i(p),
%\nonumber \\
\qquad 
f^{*}_{\alpha \beta} = k_\alpha \epsilon^{*}_\beta - 
k_\beta \epsilon^{*}_\alpha,
\nonumber
\end{eqnarray}

\noindent here $p$, $p'$, $k$ and $\epsilon$ are the 4-vectors of
the momenta of the initial and the final neutrinos and the photon, 
and the photon polarization, respectively; $m_l$ is the mass of the
intermediate charged lepton $l$; ${\cal K}_{il}$ is the unitary
lepton mixing matrix of Kobayashi-Maskava type;
$F_{\mu \nu}$ and 
$\tilde{F}_{\mu \nu} = \epsilon_{\mu \nu \alpha \beta} \,
F_{\alpha \beta} / 2$ are the tensor and the dual tensor of the constant
crossed field;
$e > 0$  is the elementary charge;  $G_F$ is the Fermi constant.
In equation~(\ref{eq:Amp2}) 
$J_{i, \sigma}(\eta_f)$ ($i = 1,2,3, \, \sigma = \pm 1, \, f = l, q$)
are the integrals of the Hardy-Stokes function $f(\eta_f)$:

\begin{eqnarray}
J_1(\eta_f) & = & \int^{1}_{0} dt \, \eta_f f(\eta_f),  
\nonumber \\
J_ {2,\sigma}(\eta_f)  & = & \int^{1}_{0} dt \, (1 + \sigma  t^2) \,
\eta_f f(\eta_f),
\nonumber \\
J _ {3,\sigma}(\eta_f) & = & \int^{1}_{0} dt \, (1 - t^2) \, 
(3 + \sigma t) \eta^2_f {df(\eta_f) \over d\eta_f},
\label{eq:Int2} \\*[.5\baselineskip]
f(\eta_f) & = & i \; \int^{\infty }_{0} du \exp \left [ -i \,
(\eta_f u + {u^3 \over 3} ) \right ], 
\nonumber \\
\eta_f &\equiv  & \bigg\{{ 16 m_f ^6 \over e_f^2 (kFFk)} 
\cdot {1\over (1 - t^2)^2} \bigg \}^{1/3}.
\nonumber
\end{eqnarray}

\noindent The functions $J_{i, \sigma}(\eta_f)$~(\ref{eq:Int2}) 
are presented in the general form, as the amplitude $\Delta M^{(X)}$
we shall analyze below can be written in terms of the integrals
$J_{i, \sigma}(\eta_q)$.

Let us discuss more detail the field-induced leptoquark contribution
$\Delta M^{(X)}$ to the amplitude $\Delta M$.
In the second order of the perturbation theory, the expression
for the matrix element  $\Delta M^{(X)}$  of the decay corresponding to 
Fig.1 with $f = q$ has the form:

\begin{eqnarray}
{\Delta M^{(X)}} & = & - 6 i \pi \cdot \frac{ e_q \alpha_s(M_X)}{M_X^2} 
\sum \limits_{q = u,c,t} {\cal U}_{iq} {\cal U}^*_{j q} 
{\varepsilon}^*_{\beta}(k)
 \,\int d^4 X \, e^{\mbox{\normalsize $ - i X k $}} 
\nonumber \\
& \times &  \, \Bigg{\{ } \, ( \bar \nu_j \nu_i ) 
\,Sp\, [ S (- X ) S ( X ) {\gamma}_{\beta} ] 
\nonumber \\
& - & ( \bar \nu_j \gamma_5 \nu_i ) 
\,Sp\, [ S (- X ) \gamma_5  S ( X ) {\gamma}_{\beta} ] 
\label{eq:M1} \\
& - & \frac {1}{2}\, ( \bar \nu_j \gamma_{\alpha} \nu_i ) 
\,Sp\, [ S (- X ) \gamma_{\alpha}  S ( X ) {\gamma}_{\beta} ] 
\nonumber \\
& - & \frac {1}{2}\, ( \bar \nu_j \gamma_{\alpha} \gamma_5 \nu_i ) 
\,Sp\, [ S (- X ) \gamma_{\alpha} \gamma_5  S (- X ) 
{\gamma}_{\beta} ] 
 \, \Bigg{\} } \; ,
\nonumber 
\end{eqnarray}

\noindent where $e_q = e Q_q$, $Q_q$ is a relative quark charge
 in the loop. 

The propagator of the quark $S (X)$ in
the crossed field in the proper time formalism~\cite{SCH} 
can be represented in the following form:

\begin{eqnarray}
S (X) & = & - {i \over 16 \pi^2} \int\limits_0^\infty
{ds \over s^2} \bigg [ {1 \over 2s} (X \gamma) +
{i e_q \over 2} (X \tilde F \gamma) \gamma_5 
\label{eq:S1} \\
& - & {s e^2_q \over 3} (X F F \gamma)
%\nonumber \\
 +  m_q - {s m_q e_q \over 2} (\gamma F \gamma) \bigg ]
\nonumber \\
& \times &  \exp \left ( - i \left [ m^2_q s + {1 \over 4 s }X^2  + 
{s e^2_q \over 12} (XFFX)  \right ] \right ),
\nonumber
\end{eqnarray}

\noindent where $X_\mu = (x-y)_\mu$; 
$\gamma_\mu$, $\gamma_5$
are  Dirac $\gamma$-matrices (where $\gamma^5 = - i \gamma^0 
\gamma^1 \gamma^2 \gamma^3$),
$m_q$ is the mass of the $up$-quark in the loop.

The integration of the amplitude~(\ref{eq:M1}) 
with respect to $X$ can be redused to the Gaussian integrals:

\begin{eqnarray}
I & = & \int d^4X \,
e^{\mbox{\large $- i \left ( \frac{1}{4} XGX + kX \right )$}}
= - (4 \pi )^2 (\det G)^{- 1/2}
e^{\mbox{\large $(i k G^{-1} k)$}},
\nonumber \\
I_\mu & = & \int d^4X \, X_\mu \,
e^{\mbox{\large $- i \left ( {1 \over 4} XGX + kX \right )$}}
= i {\partial I \over \partial k^\mu},  
\label{eq:SG} \\
I_{\mu \nu} & = & \int d^4X \, X_\mu X_\nu
e^{\mbox{\large $- i \left ( {1 \over 4} XGX + kX \right )$}}
= - {\partial^2 I \over \partial k^\mu \partial k^\nu}.
\nonumber \\
G_{\alpha\beta} & = & {s + \tau \over s \tau} g_{\alpha\beta}
 +  ( s + \tau ) {e_f^2 \over 3} \,( F F )_{\alpha\beta},
\nonumber 
\end{eqnarray}

\noindent where $s$ and $\tau$ are the proper time variables.
With ~(\ref{eq:S1}) and ~(\ref{eq:SG}) 
the invariant amplitude~(\ref{eq:M1}) can be presented in the form:

\begin{eqnarray}
{\Delta M^{(X)}} & = & 3Q_q^2 \,\, \frac{ \alpha \alpha_s(M_X)}{M_X^2} 
\sum \limits_q {\cal U}_{iq} {\cal U}^*_{j q}
\nonumber \\
& \times &
\Bigg \{  \, - ( \bar \nu_j \nu_i ) 
 {i (F f^*) \over 2 m_q} J_{2,1}(\eta_q) +
 ( \bar \nu_j \gamma_5 \nu_i ) {(\tilde F f^*) \over m_q} J_1(\eta_q) 
\label{eq:Am2}  \\
& + &
 ( \bar \nu_j \gamma_\alpha \nu_i ) 
 {e_q \over 24 m_q ^4} \left ( ( \tilde F f^* )  (k \tilde F )_\alpha 
J_{3,-1}(\eta_q) 
 - { 1 \over 2}  (\tilde F \tilde f^*)  (kF)_\alpha J_{3,1}(\eta_q)
\right )  \, 
\nonumber \\
& + & ( \bar \nu_j \gamma_\alpha \gamma_5 \nu_i ) 
(\tilde F f^*) \left ( { k_\alpha \over 8 m_q^2 } J_{ 2,-1 }(\eta_q) +
 {(kFF)_\alpha \over ( kFFk)} J_1(\eta_q)  \, \right ) \Bigg \}.
\nonumber
\end{eqnarray}

\noindent As can be readily checked, the amplitude~(\ref{eq:Am2})
is evidently gauge invariant, as it is expressed in terms of the
tensors of the external field $F_{\mu \nu}$ and the photon
field $f_{\mu \nu}$. 
We note that this expression 
does not contain the suppression factor 
$\sim m^2_q/M^2_X \ll 1 $ which is analogous to the well-known
GIM suppression factor $\sim m^2_l/M^2_W \ll 1 $
of the decay amplitude 
$\nu_i \rightarrow \nu_j \gamma$ in vacuum~\cite{Bil}.
Below we analyze a physically more interesting
case of the ultrarelativistic neutrino decay ($E_\nu \gg  m_\nu$).

\subsection{Ultrarelativistic neutrino ($E_\nu \gg m_\nu $)}

Notice that in the ultrarelativistic limit the kinematics  of
the decay $\nu_i (p) \rightarrow \nu_j (p') + \gamma (k)$ is such
that the momentum  4-vectors of the initial neutrino $p$
and the decay products $p'$  and $k$  are
almost parallel to each other. 
As the analysis shows, the contribution due to the interaction 
in the axialvector currents dominates in the
expression for amplitude (\ref{eq:Am2}) which can be 
simplified and reduced to the form:

\begin{eqnarray}
{\Delta M^{(X)}} & \simeq &  {4 \over 3} {\alpha \alpha_s ( M_X ) 
\over M_X^2} \, 
\left ( \bar \nu_j \gamma_{\alpha} \gamma_5 \nu_i \right ) \,
(\tilde{F} f ^ {*}) {(kFF)_{\alpha}\over ( kFFk)}\,< J_1(\eta_q) > 
\label{eq:Am3} \\
& \cong & {8 \sqrt 2 \over 3} \cdot 
{\alpha \alpha_s ( M_X ) \over M_X^2} \, 
( \varepsilon^* \tilde F p ) 
\left ( 1 - x + {m_j^2 \over m_i^2} ( 1 + x ) \right )^{1/2} 
< J_1(\eta_q) >, 
\nonumber \\
< J_1(\eta_q) > & = & \sum \limits_{q = u,c,t} {\cal U}_{iq} 
{\cal U}^*_{j q} \,
J_1 (m_q).
\nonumber 
\end{eqnarray}

\noindent In this case, the argument
$\eta_q$ of the Hardy-Stokes function $f(\eta_q)$  in the integral
$J_1(\eta_q)$ (see Eq~(\ref{eq:Int2})) assumes the form:

\begin{eqnarray}
\eta_q & = & 4 \left [ (1+x) (1-t^2) \left ( 1 - {m_j^2 \over m_i^2}
\right ) \chi_q \right ]^{\mbox{\normalsize - ${2 \over 3}$}},
\label{eq:Uu} \\
\chi_q^2 & = &  {e_q^{2}(p F F p) \over m_q^{6}}.
\nonumber
\end{eqnarray}

\noindent where $x=\cos \theta$, $\theta$ is the angle between vectors
$\vec p$ (the momentum of the decaying ultrarelativistic neutrino) and
$\vec {k'}$ (photon momentum in the rest system of the decaying neutrino
$\nu_i$ ). $\chi_q^2$ is the dynamic parameter which corresponds to the 
$up$-quark with the charge $e_q$ and the mass $m_q$. 
Notice  that, as $FF=F\tilde{F}=0$ in the crossed field, the dynamic 
parameter is the single field invariant, by which  the  decay  
probability is expressed.

\section{The neutrino decay probability}

The decay probability $W^{(X)}$ due to the
leptoquark contribution 
can be expressed in terms of the integral of the squared
amplitude over the variable $x$:

\begin{eqnarray}
W^{(X)} & \simeq & {1 \over 32 \pi E_\nu} \,
\left ( 1 - {m^2_j \over m^2_i} \right ) \;
\int\limits^{+1}_{-1} dx \; | \Delta M^{(X)} |^2
\nonumber \\
& = & { 4 \alpha^2 \over 9 \pi} \; {{\alpha}_s^2(M_X) 
\over E_\nu M^4_X} (p F F p) \,
\left ( 1 - {m^2_j \over m^2_i} \right )
\label{eq:WEe} \\
& \times &
\int\limits^{+1}_{-1} dx \left [ (1-x) + {m^2_j \over m^2_i} (1+x) 
\right ] \, | < J_1(\eta_q)> |^2.
\nonumber
\end{eqnarray}

\noindent Using the asymptotic expansions of the Hardy-Stokes 
functions both at large and small values of its argument 

\begin{eqnarray}
J_1(\chi_q) & = &  O(\chi_q^2), \qquad \chi_q \ll 1, \nonumber \\
J_1(\chi_q) & = & - 1 + O(\chi_q^{-2/3}), \qquad \chi_q \gg 1,
\nonumber
\end{eqnarray}

\noindent and the unitarity property of mixing matrix $U_{ij}$, 
we present here the expression (\ref{eq:WEe}) in 
the most reasonable case of the values of the dynamic parameter
$\chi_u \gg 1$, $\chi_{c, t} \ll 1$:

\begin{eqnarray}
W^{(X)} \simeq { 8 \alpha^2 \over 9 \pi} \; 
{\alpha^2_s(M_X) \over E_\nu M_X^4} \;
(p F F p) \; \left ( 1 - {m^4_j \over m^4_i} \right ) \; 
| U_{iu} U^*_{ju} |^2.
\label{eq:WEL1}
\end{eqnarray}

\noindent Here the strong hierarchy of the 
dynamic parameter $\chi_q$ 
($\chi_u : \chi_c : \chi_t = m^{- 3}_u : m^{- 3}_c : m^{- 3}_t$)
was taken into account. Because  the dynamic parameter
$\chi_q \sim (E_\nu / m_q) (F / F_q)$ is proportional to the
neutrino energy, we can see from~(\ref{eq:WEL1}) that, as the energy 
of the decaying neutrino increases, the decay probability increases
linearly. It is worth noting, that consideration
of the other limit values of the dynamic parameter
$\chi_t ( 1 - m^2_j / m^2_i ) \gg  1$
(although it isn't a physically realistic case)
shows that $W^{(X)}$ becomes finally a constant:

\begin{eqnarray}
W^{(X)} \simeq 28.8  \left (\alpha \over \pi \right )^ {3/2} \; 
{\alpha^2_s(M_X) \over E_\nu M^4_X} \;
m^3_t \sqrt{p F F p} \; | U_{it} U^*_{jt} |^2,
\label{eq:WEL2}
\end{eqnarray}

It is of interest to compare the contributions 
to the decay probability due to the $W$-boson and the leptoquark $X$
for the following values of the dynamic parameters
$\chi_u, \chi_e  \gg  1, \, \chi_{c, t}, \chi_{\mu, \tau} \ll 1$:

\begin{eqnarray}
{W^{(X)} \over W^{(W)}} & \simeq &  {16 \over 9} \; \sin^4 \Theta_W \;
\left ( {\alpha_s(M_X) \over \alpha} \right )^2 \;
\left ( {M_W \over M_X} \right )^4 \;
\frac{| {\cal U}_{iu} {\cal U}^*_{ju} |^2 } {| K_{ie} K^*_{je} |^2} 
\label{eq:WXW1} \\
& \sim & 10^{-12}\;
\left ( {100 TeV \over M_X} \right )^4 \;
\frac{| {\cal U}_{iu} {\cal U}^*_{ju} |^2 } {| K_{ie} K^*_{je} |^2} 
\nonumber
\end{eqnarray}

\noindent It follows from~(\ref{eq:WXW1}) that if 
the mixing in the lepton lector of the SM is strongly suppressed
the leptoquark contribution $W^{(X)}$ can dominate over
$W^{(W)}$.

Comparing~(\ref{eq:WEL1}) 
with the decay probability 
$\nu_i \rightarrow \nu_j \gamma$ in vacuum~\cite{Bil}:

\begin{eqnarray}
W_0 \simeq  {27 \alpha \over 32 \pi} \; {G^2_F \over 192 \pi^3} \;
\frac{m^6_i}{E_\nu}  \, \left ( {m_\tau \over m_W} \right )^4 \,
\left ( 1 + {m^2_j \over m^2_i} \right ) \,
\left ( 1 - {m^2_j \over m^2_i} \right )^3 \,
| K_{i\tau} K^*_{j\tau} |^2,    \label{eq:W0}
\end{eqnarray}

\noindent we can see the catalyzing influence of the external field
on the neutrino radiative decay 
in the $SU(4)_V \times SU(2)_L \times G_R$ Model
in case of the leptoquark contribution to the decay probability:

\begin{eqnarray}
{W^{(X)} \over W_0} & \simeq & {2^{11} \over 9} \; \sin^4 \Theta_W \;
\left ( {\alpha_s(M_X) \over \alpha} \right )^2 \;
\chi_u^2 \;
\left ( {m_u \over m_i} \right )^6 \;
\left ( {M_W \over m_\tau} \right )^4 \;
\left ( {M_W \over M_X} \right )^4 \;
\frac{| {\cal U}_{iu} {\cal U}^*_{ju} |^2 } {| K_{i\tau} K^*_{j\tau} |^2} 
%\label{eq:WXW2}
\nonumber \\
& \sim & 10^{37} \; \chi_u^2 \;
\left ( {1 eV \over m_i} \right )^6 \;
\left ( {100 TeV \over M_X} \right )^4 \;
\frac{| {\cal U}_{iu} {\cal U}^*_{ju} |^2 } {| K_{i\tau} K^*_{j\tau} |^2} 
%\nonumber 
\label{eq:WXW2}
\end{eqnarray}

\noindent Such a gigantic enhancement is that the external field
removes the main suppression caused by the
smallness of the neutrino mass.

\section{Conclusion} 

The influence of the external crossed field on the massive neutrino 
radiative decay $\nu_i \rightarrow \nu_j \gamma$ ($i \ne j$) 
was studied in the framework of the 
Minimal Quark-Lepton Symmetry of the Pati-Salam type based on the 
$SU(4)_V \times SU(2)_L \times G_R$ group. The matrix element
$\Delta M^{(X)}$ and the decay probability $W^{(X)}$ due to the
leptoquark $X$ contribution were obtained. 

1. The most substantial manifestation of the
catalyzing influence of the external field is that the field removes 
the main suppression associated with the smallness of the
mass of the decaying neutrino (in vacuum this suppression is
$W_0 \sim (m^6_\nu / E_\nu)^6$ for decay in flight).
The decay probabilities ~(\ref{eq:WEL1}) and~(\ref{eq:WEL2}) do not
depend of the specific neutrino masses, if only the threshold factor
$( 1 - m^4_j/m^4_i )$ is not close to zero.

2. In the expressions ~(\ref{eq:WEL1}) and~(\ref{eq:WEL2}) 
the suppression factor 
$\sim m^2_q/M^2_X \ll 1 $ which is analogous to the well-known
GIM suppression factor $\sim m^2_l/M^2_W \ll 1 $ of the decay  
$\nu_i \rightarrow \nu_j \gamma$ in vacuum~\cite{Bil} is absent.

3. The leptoquark contribution to the decay 
probability can dominate over the $W$-boson one 
(see eq.~(\ref{eq:WXW1}))
in the case of the strong suppression of the lepton mixing
in the framework of the SM:

\begin{eqnarray}
| {\cal U}_{iq} {\cal U}^*_{jq} | >  10^{6}\; 
\left ( {M_X \over 100 TeV} \right )^4 \;
| K_{il} K^*_{jl} |,
\nonumber
\end{eqnarray}

\vspace*{5mm}

{\bf Acknowledgments} 

\vspace*{5mm}

The work of Mikheev N.V. was supported by a Grant N d104
by International Soros Science Education Program.
The work of Vassilevskaya L.A. was supported by a fellowship
of INTAS Grant 93-2492-ext and was carried out within the research
program of International Center for Fundamental Physics in Moscow.

\end{document}